%% file: SwitchingWeakChimeras_final.tex
\documentclass[aps, pre, twocolumn, groupedaddress, showpacs, floatfix,10pt]{revtex4-1}

\usepackage{amsmath, amssymb}
\usepackage[english]{babel}
\usepackage[latin1]{inputenc}
\usepackage[all]{xy}
\usepackage{graphicx}
\usepackage{tikz}
\usetikzlibrary{matrix}
\usepackage{verbatim}
\usepackage{enumerate}
\usepackage{url}
\usepackage{subfigure}
\usepackage{bm}
\usepackage{times}
\usepackage{xcolor}

\include{commands}

\newcommand{\Xsk}{X_{\sigma,k}}
\newcommand{\Xso}{X_{\sigma,1}}
\newcommand{\Xst}{X_{\sigma,2}}

\newcommand{\CS}{K}

\newcommand{\twocol}[1]{#1}
\newcommand{\half}{}

\begin{document}

\preprint{}


\title{Heteroclinic switching between chimeras}
\author{Christian Bick}%

\affiliation{%
\twocol{\mbox}
{Oxford Centre for Industrial and Applied Mathematics, Mathematical Institute, University of Oxford, OX2 6GG, UK}\\
\twocol{\mbox}
{Centre for Systems Dynamics and Control and Department of Mathematics, University of Exeter, EX4~4QF, UK}
}
\date{\today}


\begin{abstract}
Functional oscillator networks, such as neuronal networks in the brain, exhibit switching between metastable states involving many oscillators. We give exact results how such global dynamics can arise in paradigmatic phase oscillator networks: higher-order network interactions give rise to metastable chimeras---localized frequency synchrony patterns---which are joined by heteroclinic connections. Moreover, we illuminate the mechanisms that underly the switching dynamics in these experimentally accessible networks.
\end{abstract}

\maketitle

\noindent
Networks of (almost) identical nonlinear oscillators give rise to fascinating collective dynamics where populations of localized oscillators exhibit distinct frequencies and levels of phase synchronization~\cite{Litwin-Kumar2012, Martens2013}. In neuronal networks, the location of such localized frequency synchrony patterns can encode information~\cite{Hubel1959, Tognoli2014, Bick2014a}. Thus, sequential switching between distinct localized dynamics has been associated with neural computation~\cite{AshwinTimme2005, Rabinovich2008a, Britz2010, Ashwin2015}; sequential dynamics in the hippocampus in the absence of external input~\cite{Wilson1994} are a striking example. Most efforts to understand switching dynamics between localized frequency synchrony patterns rely on averaged models which neglect the contributions of individual oscillators to the network dynamics~\cite{Kiebel2009, Bick2009a, Rabinovich2015, Horchler2015, Schaub2015} or are statistical~\cite{Wildie2012}. For finite networks, however, the dynamics of individual oscillators cannot be neglected.

In this article we give explicit results for the emergence of switching between synchrony patterns that are characterized by localized frequency synchrony---commonly known as (weak) chimeras~\cite{Panaggio2015, Scholl2016}---in phase oscillator networks with higher-order interactions. More precisely, we prove the existence of saddle weak chimeras which are joined by heteroclinic connections; nearby trajectories exhibit sequential switching of localized frequency synchrony. Our results directly relate two distinct dynamic phenomena, heteroclinic switching and chimeras, and thus give a number of insights into the global dynamics of oscillator networks. First, they elucidate how network topology and the functional form of the oscillator coupling facilitate switching dynamics: the heteroclinic structures arise through an interplay of higher harmonics in the phase coupling function and interaction terms which depend on the phase differences of more than two oscillators (nonpairwise interaction). Although such generalized forms of network coupling arise naturally in phase reductions of generically coupled limit cycle oscillator networks~\cite{Ashwin2015a}, they are neglected in classical Kuramoto-type networks~\cite{Acebron2005, Rodrigues2016}. Hence, our results emphasize how higher-order interaction terms can shape the phase dynamics of many physical systems, from oscillator networks~\cite{Rosenblum2007, Bick2016b, Tanaka2011a} to ecological systems~\cite{Levine2017}. Second, switching between metastable chimeras is an explicit dynamical mechanism how networks of neural oscillators may encode sequential information and give rise to dynamics similar to hippocampal replay. Third, we provide a theoretical foundation to understand self-organized switching between chimeras that was recently observed in numerical simulations~\cite{Haugland2015, Maistrenko2016}. Finally, relating heteroclinic switching and chimeras opens up a range of questions; for example, whether any given itinerary can be realized as a heteroclinic structure between chimeras.

In the following, we consider networks of~$\maxpop$ populations of~$\maxdim$ phase oscillators. Let $\theta_{\sigma,k}\in\Tor := \R/2\pi\Z$ denote the phase of oscillator~$k$ in population~$\sigma$. Write $\theta=(\theta_{1}, \dotsc, \theta_{\maxpop})\in\Tormn$ where $\theta_\sigma = (\theta_{\sigma, 1}, \dotsc, \theta_{\sigma, \maxdim})\in\Torn$ is the state of population~$\sigma$. The set $\Sync := \tset{(\phi_1, \dotsc, \phi_\maxdim)\in\Torn}{\phi_k=\phi_{k+1}}$ corresponds to  phase synchrony and $\Splay := \tset{(\phi_1, \dotsc, \phi_\maxdim)\in\Torn}{\phi_{k+1}=\phi_{k}+\frac{2\pi}{\maxdim}}$ denotes the splay phase where phases are distributed uniformly on the circle. Following~\cite{Martens2010c} we use the shorthand notation
\begin{subequations}\label{eq:SyncSplay}
\begin{align}
\theta_1\dotsb\theta_{\sigma-1}\Sp\theta_{\sigma+1}\dotsb\theta_{\maxpop} &= \lset{\theta\in\Tormn}{\theta_\sigma\in\Sync}\\
\theta_1\dotsb\theta_{\sigma-1}\Dp\theta_{\sigma+1}\dotsb\theta_{\maxpop} &= \lset{\theta\in\Tormn}{\theta_\sigma\in\Splay}
\end{align}
\end{subequations}
to indicate that population~$\sigma$ is phase synchronized or in splay phase. Hence, $\Sp\dotsb\Sp$ ($\maxpop$~times) is the set of cluster states and $\Dp\dotsb\Dp$ is the set where all populations are in splay phase. Given a dynamical system on~$\Tormn$ and a trajectory~$\theta(t)$ with initial condition $\theta(0)=\theta^0$, define the asymptotic average angular frequency $\Omega_{\sigma,k}(\theta^0):=\lim_{t\to\infty}\frac{1}{t}\theta_{\sigma,k}(t)$. The characterizing feature of a \emph{weak chimera} as an invariant set~$A\subset\Tormn$ is \emph{localized frequency synchrony}: for all $\theta^0\in A$ we have oscillators $(\sigma,k)$, $(\tau,j)$, $(\rho,\ell)$ such that $\Omega_{\sigma,k}(\theta^0)=\Omega_{\tau,j}(\theta^0)\neq\Omega_{\rho,\ell}(\theta^0)$; see also~\cite{Ashwin2014a, Bick2015c, Bick2015d}.

\newcommand{\figonepanell}[1]{\raisebox{3.7cm}{(#1)}\hspace{-10pt}}

\begin{figure*}
\includegraphics[width=\linewidth]{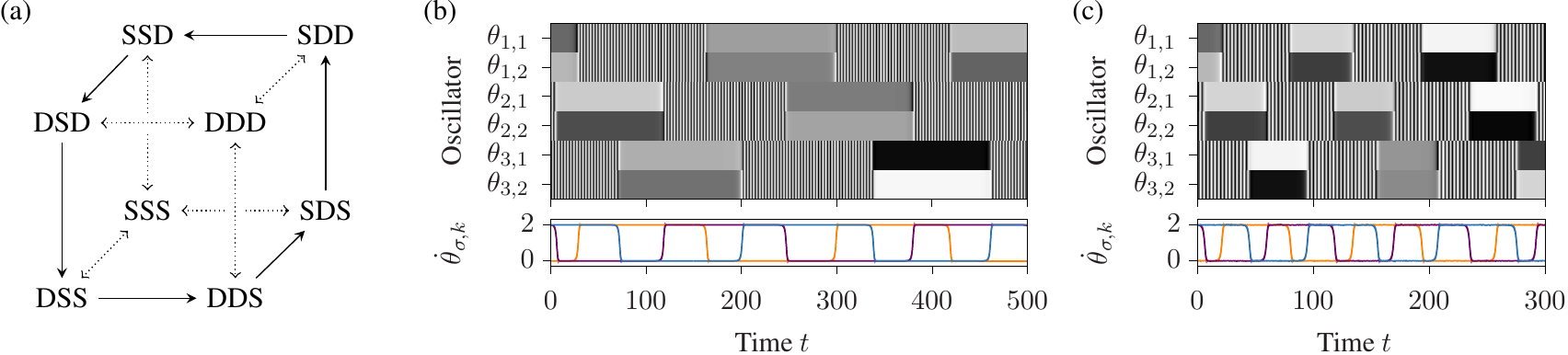}
\caption{\label{fig:HetCycle3x2}Heteroclinic networks appear in networks of $\maxpop=3$ populations of $\maxdim=2$ oscillators~\eqref{eq:Dyn3x2Num} with coupling function~\eqref{eq:CplngFuncN2} and $\alpha=\frac{\pi}{2}$, $r=-0.1$, $\CS=0.1$. Panel~(a) shows the heteroclinic cycle between saddle weak chimeras (solid lines); stability is indicated by arrows. Panel~(b) shows switching of localized frequency synchrony close to the heteroclinic cycle, $\delta=0$, in the presence of noise, $\eta=10^{-7}$: the oscillators' phases (shading; black if $\theta_{\sigma,k}=\pi$, white if $\theta_{\sigma,k}=0$) and frequencies (colors; shading indicates a population's frequency range $[\min_k\dot\theta_{\sigma,k}, \max_k\dot\theta_{\sigma,k}]$, a line its average $\langle\dot\theta_{\sigma, k}\rangle_k$) are plotted over time. The frequencies of different populations synchronize and desynchronize sequentially. Panel~(c) shows irregular switching dynamics without noise, $\eta=0$, when the symmetries are broken, $\delta=0.01$. 
}
\end{figure*}

\textit{Heteroclinic cycles in small networks.---}%
Consider a network of $\maxpop=3$ populations of $\maxdim=2$ identical phase oscillators where the interaction within populations is pairwise and determined by the coupling function
\begin{equation}\label{eq:CplngFuncN2}
g(\vth) = \sin(\vth+\alpha)+r\sin(2(\vth+\alpha))
\end{equation}
parametrized by $\alpha, r\in\R$, whereas different populations interact at coupling strength~$\CS$ through the \textbf{s}inusoidal \textbf{n}on\textbf{p}airwise interaction function
\begin{equation}\label{eq:NPCN2}
\begin{split}
\snd(\phi, \vartheta; \theta_\tau) &= \cos(\theta_{\tau,1}-\theta_{\tau,2}+\phi-\vartheta+\alpha)
\twocol{\\&\qquad}
+\cos(\theta_{\tau,2}-\theta_{\tau,1}+\phi-\vartheta+\alpha).
\end{split}
\end{equation}
More specifically, the dynamics of population~$\sigma\in\sset{1,2,3}$ is given by
{\allowdisplaybreaks
\begin{subequations}\label{eq:Dyn3x2}
\begin{align}
\begin{split}
\dot\theta_{\sigma,1} &= \omega+g(\theta_{\sigma,2}-\theta_{\sigma,1})-\CS\snd(\theta_{\sigma,2},\theta_{\sigma,1};\theta_{\sigma-1})
\twocol{\\&\qquad\qquad}
+\CS\snd(\theta_{\sigma,2},\theta_{\sigma,1}; \theta_{\sigma+1})=:\Xso(\theta),
\end{split}
\\
\begin{split}
\dot\theta_{\sigma,2} &=  \omega+g(\theta_{\sigma,1}-\theta_{\sigma,2})-\CS\snd(\theta_{\sigma,1},\theta_{\sigma,2};\theta_{\sigma-1})
\twocol{\\&\qquad\qquad}
+\CS\snd(\theta_{\sigma,1},\theta_{\sigma,2}; \theta_{\sigma+1})=:\Xst(\theta),
\end{split}
\end{align}
\end{subequations}
}%
where $\omega$ is the oscillators' intrinsic frequency~\footnote{Note that we can set~$\omega$ to any value without loss of generality by going in a suitable co-rotating reference frame. In the figures we set $\omega=-\sum_{j=1}^{N-1}g(2\pi j/N)$ so that for $\CS=0$ the splay configuration appears stationary.} and indices are taken modulo~$\maxpop$.

The coupling induces symmetries of the oscillator network. For each of the $\maxpop$ populations, let $\Tor$ act by shifting all phases of that population by a common constant and let the symmetric group~$\Sn$ permute its~$\maxdim$ oscillators. Suppose that $\Zm:=\Z/\maxpop\Z$ permutes populations cyclically. The equations of motion~\eqref{eq:Dyn3x2} are invariant under the group of transformations $(\Sn\times\Tor)^\maxpop\rtimes \Zm$ of~$\Tormn$. The semidirect product~``$\rtimes$'' indicates that actions do not necessarily commute~\cite{Ashwin1992}. These symmetries induce invariant subspaces~\cite{Golubitsky2002}: in particular $\SSS$, $\DDD$ as well as $\DSS$, $\DDS$ and their images under permutations of populations are dynamically invariant.

We can now give conditions for~\eqref{eq:Dyn3x2} to have the heteroclinic cycle depicted in Fig.~\ref{fig:HetCycle3x2}(a) between saddle weak chimeras $\DSS$, $\DDS$ and their symmetric counterparts. Because of symmetry, it suffices to consider $\DSS$, $\DDS$. We proceed in three steps. First, we want $\DSS$, $\DDS$ to be weak chimeras. Second, we give conditions for the invariant sets to be saddles. Third, we show that they are connected by heteroclinic orbits. Here we focus on the case of $\alpha=\frac{\pi}{2}$ and refer to~\cite{Bick2018} for more generality and a proof that there is in fact an open set of parameters~$(\alpha, \CS, r)$ for which this heteroclinic cycle between weak chimeras exists.

First, for $\DSS$, $\DDS$ to be weak chimeras, we calculate the frequencies~$\Omega_{\sigma,k}$ for~\eqref{eq:Dyn3x2}. For $\CS=0$ we have $\Omega_{1,k}(\theta^0)=\omega+1$ for $\theta^0\in \Sp\theta_2\theta_3$ and $\Omega_{2,k}(\theta^0) = \omega-1$ for $\theta^0\in \theta_1\Dp\theta_3$. In other words, without coupling between populations, the frequency difference between a synchronized and an anti-phase population is $\abs{(\omega+1)-(\omega-1)}=2$. With coupling, $\CS>0$, the maximal change in frequency difference is proportional to~$\CS$. Specifically, using the triangle inequality in~\eqref{eq:Dyn3x2} yields that ${\Omega_{1,k}(\theta^0)\neq\Omega_{2,k}(\theta^0)}$ for $\theta^0\in\Sp\Dp\theta_3$ if $2-8\CS>0$. At the same time, ${\Omega_{\sigma,k}(\theta^0)=\Omega_{\sigma,j}(\theta^0)}$ for all $\theta^0\in\Tormn$ with $\theta^0_\sigma\in\Sync,\Splay$. Hence, $\DSS$, $\DDS$ are weak chimeras for~\eqref{eq:Dyn3x2} on~$\Tormn$ if $2\CS<\frac{1}{2}$.

Second, we need $\DSS, \DDS$ to be saddle invariant sets. Reduce the phase-shift symmetries by rewriting~\eqref{eq:Dyn3x2} in terms of phase differences $\psi_{\sigma,k} := \theta_{\sigma, k+1} - \theta_{\sigma, 1}$, $k=1, \dotsc, \maxdim-1$. (Consequently, we may replace all~$\theta$ by the phase differences~$\psi$ in~\eqref{eq:SyncSplay}.) Since $\maxdim=2$ here, $\psi_\sigma=\psi_{\sigma,1}$ determines the state of population~$\sigma$ and the effective dynamics of~\eqref{eq:Dyn3x2} are three-dimensional. In the reduced system~$\DSS=(\pi,0,0)$, $\DDS=(\pi,\pi,0)$ are equilibria. Linearizing at~$\DSS$ yields eigenvalues 
$\lambda^\DSS_1 = 
4r$,
$\lambda^\DSS_2 = 
8\CS+4r$,
$\lambda^\DSS_3 = 
-8\CS+4r$
that correspond to linear stability of the first, second, and third population, respectively. Similarly, for~$\DDS$ we obtain the eigenvalues
$\lambda^\DDS_1=8\CS+4r$,
$\lambda^\DDS_2=-8\CS+4r$,
$\lambda^\DDS_3=4r$.
Observe that if $0<-r<2\CS$ we have $\lambda^\DSS_1=\lambda^\DDS_3<0$, $\lambda^\DSS_2=\lambda^\DDS_1>0$, $\lambda^\DSS_3=\lambda^\DDS_2<0$ and thus $\DSS, \DDS$ are saddle invariant sets with two-dimensional stable and one-dimensional unstable manifolds.

Third, we obtain conditions for heteroclinic connections between~$\DSS, \DDS$ given their stability above. Observe that $\lambda^\DSS_2>0$, $\lambda^\DDS_2<0$ implies that the unstable manifold of~$\DSS$ and the stable manifold of~$\DDS$ both intersect the invariant subspace~$\Dp\psi_2\Sp$ on which the dynamics reduce to $\dot\psi_2 = \sin(\psi_2)(8\CS-4r\cos(\psi_2))$. Thus, if $-r<2\CS$ there are no equilibria other than $\psi_2\in\sset{0,\pi}$ (these are $\DSS$ and $\DDS$) in $\Dp\psi_2\Sp$ and we have a heteroclinic connection. Indeed, we get the same condition for there to be no additional equilibria in~$\psi_1\Dp\Sp$. To summarize, for $\alpha=\frac{\pi}{2}$ the heteroclinic cycle sketched in~Fig.~\ref{fig:HetCycle3x2}(a) exists if $0<-r<2\CS<\frac{1}{2}$. Moreover, one can show by evaluating the saddle values that for $\CS<-r$ the cycle is expected to attract nearby initial conditions~\cite{Bick2018}.

The switching dynamics between weak chimeras persists when the particular nonpairwise coupling scheme  of~\eqref{eq:Dyn3x2} is broken. With noise given by a Wiener process~$W_{\sigma,k}$ (Brownian motion) and a symmetry breaking coupling term $S_{\sigma, k}(\theta) = \Delta\omega_{\sigma,k}+\frac{1}{\maxpop\maxdim}\sum_{\tau=1}^\maxpop\sum_{j=1}^\maxdim\sin(\theta_{\tau,j}-\theta_{\sigma,k})$ with normally distributed frequency deviations $\Delta\omega_{\sigma,k}$ (mean zero and variance one), we integrated the system
\begin{equation}\label{eq:Dyn3x2Num}
\dot\theta_{\sigma,k} = \Xsk(\theta) + \delta S_{\sigma, k}(\theta) + \eta W_{\sigma,k}
\end{equation}
numerically in XPP~\cite{Ermentrout2002} where~$\Xsk$ as in~\eqref{eq:Dyn3x2}. For $\eta>0$, $\delta=0$ we obtain heteroclinic switching where transition times scale with the noise amplitude~$\eta$ as expected~\cite{Stone1990}; cf.~Fig.~\ref{fig:HetCycle3x2}(b). Setting $\delta>0$ breaks all symmetries to a single phase-shift symmetry acting as a common phase shift for all oscillators. Although this breaks the invariant subspaces containing the heteroclinic connections, we still obtain sequential dynamics prescribed by the heteroclinic network as shown in Fig.~\ref{fig:HetCycle3x2}(c).

\textit{Order parameter dependent coupling induces switching.---}%
The dynamical mechanism which leads to heteroclinic cycles in~\eqref{eq:Dyn3x2} can be best understood if the oscillator network is seen as individual populations coupled through their mean fields. Let $i=\sqrt{-1}$. The absolute value of the Kuramoto order parameter $R_\sigma := R(\theta_\sigma) = \tabs{\frac{1}{\maxdim}\sum_{j=1}^{\maxdim}\exp(i\theta_{\sigma, j})}$ gives information about synchronization: $\theta_\sigma\in\Sync$ iff $R(\theta_\sigma)=1$ and $\theta_\sigma\in\Splay$ implies $R(\theta_\sigma)=0$. For $a\in\N$ let
\begin{equation}\label{eq:CplngFunc}
g(\vth) = \sin(\vth+\alpha)+r\sin(a(\vth+\alpha))
\end{equation}
generalize the coupling function~\eqref{eq:CplngFuncN2}. Now consider a system of~$\maxpop$ populations of~$\maxdim$ phase oscillators each where the dynamics of oscillator~$k$ in population~$\sigma$ are given by
\begin{equation}\label{eq:DynMxNFull}
\dot\theta_{\sigma,k} = \omega+\frac{1}{\maxdim}\sum_{j\neq k} g(\theta_{\sigma,j}-\theta_{\sigma,k}+\Delta\alpha_\sigma)
\end{equation}
and~$\Delta\alpha_{\sigma}$ modulates the phase-shift~$\alpha$ of the coupling function~\eqref{eq:CplngFunc}. If $r=0$ then either full synchrony~$\Sync$ or the phase configurations with $R_\sigma=0$ are globally attracting for~\eqref{eq:DynMxNFull} depending on the value of~$\alpha+\Delta\alpha_\sigma$~\footnote{Except for some set of initial conditions of zero Lebesgue measure.}. In particular, the global attractors swap stability at $\alpha+\Delta\alpha_\sigma=\pm{\frac{\pi}{2}}$. 
Hence, for $r=0$ and $\alpha\approx\frac{\pi}{2}$ the order parameter-dependent modulation of~$\Delta\alpha_{\sigma}$ by
\begin{equation}\label{eq:OPdBP}
\Delta\alpha_{\sigma} = \CS((1-R_{\sigma-1}^2) - (1-R_{\sigma+1}^2)),
\end{equation}
$0<K\lessapprox\frac{\pi}{2}$, yields a mechanism for sequential synchronization:
If population $\sigma-1$ is synchronized ($R_{\sigma-1}=1$) and population $\sigma+1$ is in splay phase ($R_{\sigma+1}=0$) then~$\Sync$ is asymptotically stable for population~$\sigma$. Conversely, if $R_{\sigma+1}=1$ and $R_{\sigma-1}=0$ then~$R_\sigma=0$ is asymptotically stable for population~$\sigma$. Whereas the system is degenerate for $R_{\sigma-1}=R_{\sigma+1}$ if $\alpha=\frac{\pi}{2}$ and $r=0$, an appropriate choice of~$a$ and $r\neq 0$ to induce bistability of~$\Sync$ and~$\Splay$ will resolve the degeneracy below.

A network with nonpairwise coupling approximates the system~\eqref{eq:DynMxNFull} with state-dependent phase shift~\eqref{eq:OPdBP}. We have
\begin{equation}\label{eq:Subst1}
\begin{split}
g(\vth +\Delta\alpha_\sigma) &= 
g(\vth)+\CS(R_{\sigma+1}^2-R_{\sigma-1}^2)\cos(\vth+\alpha)
\twocol{\\&\qquad}
+ O\!\left(\CS^2\right)+ O\!\left(\CS r\right).
\end{split}
\end{equation}
Generalizing~\eqref{eq:NPCN2}, define the \textbf{s}inusoidal \textbf{n}on\textbf{p}airwise \textbf{s}caled interaction function
\[
\snx(\phi, \vth; \theta_\tau) = \frac{1}{\maxdim^2}\sum_{p,q=1}^\maxdim\cos(\theta_{\tau, p}-\theta_{\tau,q} + \phi-\vth+\alpha).
\]
Note that
$R_\tau^2 = \frac{1}{\maxdim^2}\sum_{p,q=1}^\maxdim \cos(\theta_{\tau,p}-\theta_{\tau,q})$
which implies
\begin{align}\label{eq:Subst2}
R_\tau^2\cos\left(\theta_{\sigma, j}-\theta_{\sigma, k}+\alpha\right) &= \snx(\theta_{\sigma, j}, \theta_{\sigma, k}; \theta_\tau).
\end{align}
Substituting~\eqref{eq:Subst1} and~\eqref{eq:Subst2} into~\eqref{eq:DynMxNFull} and dropping the $O\big(\CS^2\big)$, $O\!\left(\CS r\right)$ terms yields the phase dynamics
\begin{align}\label{eq:DynMxN}
\twocol{\nonumber}
\dot\theta_{\sigma,k} &= \omega+\frac{1}{\maxdim}\sum_{j\neq k}\Big(g(\theta_{\sigma, j}-\theta_{\sigma, k})
-\CS\snx(\theta_{\sigma, j}, \theta_{\sigma, k}; \theta_{\sigma-1})
\twocol{\\&\qquad\quad}
+\CS\snx(\theta_{\sigma, j}, \theta_{\sigma, k}; \theta_{\sigma+1}) 
\Big)
 =: \Xsk(\theta)
\end{align}
as an approximation of~\eqref{eq:DynMxNFull}. Note that for $\maxpop=3$, $\maxdim=2$, the system~\eqref{eq:Dyn3x2} with coupling function~\eqref{eq:CplngFuncN2} is---up to rescaling of~$\CS$ and time---exactly this approximation~\eqref{eq:DynMxN} with~\eqref{eq:CplngFunc} and harmonic~$a=2$ that yields hyperbolic saddles.

\begin{figure}
\includegraphics[width=\half\linewidth]{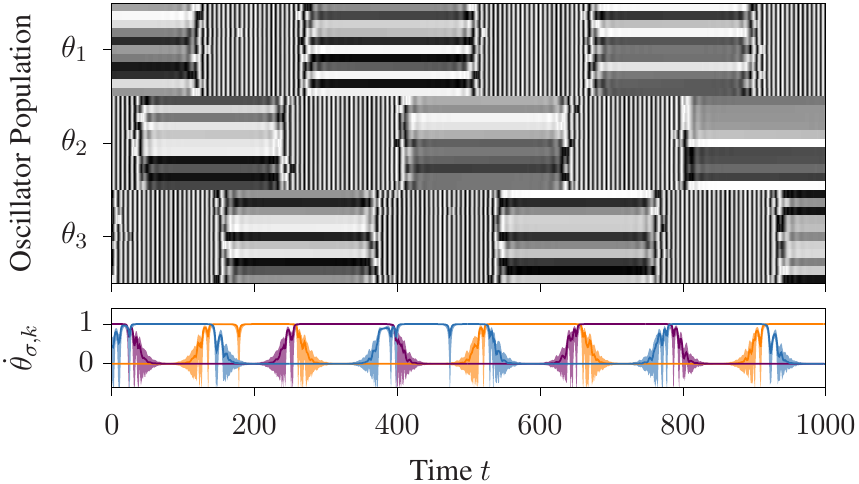}
\caption{\label{fig:HetSwitch3xHighD}%
Switching between localized frequency synchrony is observed in networks of $\maxpop=3$ populations of $\maxdim=11$ oscillators with dynamics~\eqref{eq:Dyn3x2Num} and vector field~\eqref{eq:DynMxN}. As in Fig.~\ref{fig:HetCycle3x2} the evolution of phases and order parameters the oscillators populations synchronize in frequency sequentially. Coupling for $\CS = 0.2$ is given by~\eqref{eq:CplngFunc} with $a=22$, $r = -0.001$, $\alpha=\frac{\pi}{2}$, symmetry breaking $\delta = 0.001$ and no noise, $\eta=0$.
}
\end{figure}

\textit{Switching dynamics for larger networks.---}%
The derivation of the nonpairwise coupling suggests a general mechanism to obtain switching dynamics in systems with population sizes $\maxdim>2$. Indeed, we obtain sequential switching dynamics for example for $\maxpop=3$, $\maxdim=11$: integrating~\eqref{eq:Dyn3x2Num} with~$\Xsk$ as in~\eqref{eq:DynMxN} yields sequential switching  even when the system symmetries are broken, $\delta>0$; cf.~Fig.~\ref{fig:HetSwitch3xHighD}. Note that the transitions now take place along high-dimensional invariant subspaces.


\newcommand{\figthreepanelo}[1]{\raisebox{4.1cm}{(#1)}\hspace{-25pt}}
\newcommand{\figthreepanelt}[1]{\raisebox{4.1cm}{(#1)}\hspace{-10pt}}

\begin{figure}
\includegraphics[width=\half\linewidth]{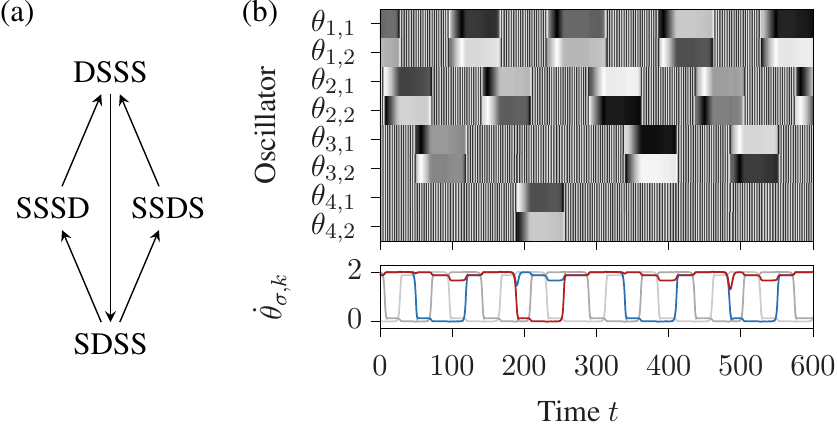}
\caption{\label{fig:HetNet4x2}The network of $\maxpop=4$ populations of $\maxdim=2$ oscillators with dynamics~\eqref{eq:Dyn4x2} shows noise induced random switching from~$\SDSS$ to either~$\SSDS$ or~$\SSSD$. This relates to a Kirk--Silber type network sketched in Panel~(a). Panel~(b) depicts evolution of phases and frequencies (populations 3 and 4 are highlighted in color) for coupling~\eqref{eq:CplngFuncN2} with $r = -0.1$, $\alpha=\frac{\pi}{2}$, $\CS = 0.35$, and $\eta=10^{-4}$.
}
\end{figure}

\textit{From heteroclinic cycles to networks.---}%
Generalizing the order parameter-dependent coupling~\eqref{eq:OPdBP} for the dynamics~\eqref{eq:DynMxNFull} leads to switching similar to those observed for the Kirk--Silber heteroclinic network~\cite{Kirk1994} which contains more than one cycle; cf.~Fig.~\ref{fig:HetNet4x2}(a). 
Similar to~\eqref{eq:OPdBP}, set
\begin{subequations}\label{eq:PhaseShift4x2}
\begin{align}
\Delta\alpha_1&=-\CS(1-R_{2}^2)+\CS(1-R_{3}^2)+\CS(1-R_{4}^2),\\
\Delta\alpha_2&=\CS(1-R_{1}^2)-\CS(1-R_{3}^2)-\CS(1-R_{4}^2),\\
\Delta\alpha_3&=-\CS(1-R_{1}^2)+\CS(1-R_{2}^2)-\CS(1-R_{4}^2),\\
\Delta\alpha_4&=-\CS(1-R_{1}^2)+\CS(1-R_{2}^2)-\CS(1-R_{3}^2).
\end{align}
\end{subequations}
Consider $\maxpop=4$ populations of $\maxdim=2$ oscillators where oscillator~$(\sigma, k)$ evolves according to
\begin{equation}\label{eq:Dyn4x2}
\dot\theta_{\sigma,k} = \omega + g\big(\theta_{\sigma,3-k}-\theta_{\sigma,k} +\Delta\alpha_\sigma\big) + \eta W_{\sigma,k}
\end{equation}
with coupling function~$g$ as in~\eqref{eq:CplngFuncN2}. The~$\Delta\alpha_{\sigma}$ given by~\eqref{eq:PhaseShift4x2} are now chosen to allow for switching from~$\SDSS$ to either~$\SSDS$ or~$\SSSD$: if population~2 is desynchronized, $R_2=0$, and all other populations are synchronized, $R_\sigma=1, \sigma\neq 2$ then~$\Dp$ will be attracting for both populations~3 and~4 (in the limiting case $r=0$). Fig.~\ref{fig:HetNet4x2}(b) shows noise-induced switching in~\eqref{eq:Dyn4x2}. 
A full analysis of this system (and its nonpairwise approximation) is beyond the scope of this article.

\textit{Discussion---}%
Phase oscillator networks with nonpairwise coupling have surprisingly rich dynamics~\cite{Rosenblum2007, Tanaka2011a, Ashwin2015a, Bick2016b}; here, nonpairwise interaction allows to show the existence of heteroclinic connections between weak chimeras. Here nonpairwise coupling arises through a bifurcation parameter that depends on \emph{local} order parameters of different populations. By contrast, the dynamics of a network with a bifurcation parameter depending on the \emph{global} order parameter has been studied in their own right~\cite{Burylko2011} and exploited for applications~\cite{Sieber2014}. In contrast to sequential switching of phase synchrony for nonidentical oscillators~\cite{Komarov2011}, here we observe switching of localized frequency synchrony in a network of indistinguishable phase oscillators (the symmetry action is transitive). Moreover, since the system is close to bifurcation for small~$\CS$, small perturbations to the vector field allow for going from one switching sequence to another. 

Our results open up a range of questions relating both chimeras and heteroclinic networks. Are there heteroclinic cycles between saddle weak chimeras with chaotic dynamics~\cite{Bick2015c}? Is it possible to realize any heteroclinic network in a phase oscillator network where the saddles are weak chimeras, see also~\cite{Ashwin2013, Field2015}? How do the dynamics of~\eqref{eq:Dyn4x2} relate to results obtained for the Kirk--Silber network~\cite{Castro2016}?

Heteroclinic switching between localized frequency synchrony patterns is of direct relevance for real-world systems. On the one hand, note that the small networks considered here are accessible for experimental realizations: weak chimeras have recently been observed in electrochemical systems~\cite{Bick2017} with linear and quadratic interactions interactions~\cite{Kori2008}. Thus, we are interested in whether switching of localized frequency synchrony is observed these experimental setups. On the other hand, sequential switching of localized frequency synchrony may be an important aspect of functional dynamics in networks of neurons. Our results elucidate the features of network interaction (e.g., symmetries and nonpairwise interactions) and the dynamical mechanisms that facilitate switching dynamics. Thus, our insights may open up ways to restore and control functional dynamics, for example, if the network becomes pathologically synchronized.

\textit{Acknowledgements---}%
The author would like to thank M.~Field, E.~A.~Martens, O. Omel'chenko, T.~Pereira, M.~Rabinovich, M.~Wolfrum, and in particular P.~Ashwin for many helpful discussions.
This work has received funding from the People Programme (Marie Curie Actions) of the European Union's Seventh Framework Programme (FP7/2007--2013) under REA grant agreement no.~626111.


\bibliographystyle{apsrev4-1}
\def\urlprefix{}
\def\url#1{}

\bibliography{ref}

\end{document}

%% file: commands.tex
\newcommand{\maxdim}{N}
\newcommand{\maxpop}{M}

\DeclareMathOperator{\snd}{snp}
\DeclareMathOperator{\snx}{snps}

\newcommand{\Splay}{\Dp}
\newcommand{\Sync}{\Sp}

\newcommand{\vth}{\vartheta}

\newcommand{\R}{\mathbb{R}}

\newcommand{\Z}{\mathbb{Z}}
\newcommand{\N}{\mathbb{N}}

\newcommand{\Zm}{\Z_\maxpop}

\newcommand{\Tor}{\mathbf{T}}
\newcommand{\Torn}{\Tor^\maxdim}

\newcommand{\Tormn}{\Tor^{\maxpop\maxdim}}

\newcommand{\Ss}{\mathbf{S}}

\newcommand{\Sn}{\Ss_\maxdim}

\newcommand{\abs}[1]{\left|#1\right|}
\newcommand{\tabs}[1]{\big|#1\big|}

\newcommand{\lset}[2]{\left\lbrace\left. #1\;\right|\,#2\,\right\rbrace}

\newcommand{\tset}[2]{\big\lbrace #1\,\big|\;#2\big\rbrace}
\newcommand{\sset}[1]{\left\lbrace #1\right\rbrace}

\newcommand{\Sp}{\textrm{S}}
\newcommand{\Dp}{\textrm{D}}
\newcommand{\DDD}{{\Dp\Dp\Dp}}
\newcommand{\SSS}{{\Sp\Sp\Sp}}
\newcommand{\DSS}{{\Dp\Sp\Sp}}

\newcommand{\DDS}{{\Dp\Dp\Sp}}

\newcommand{\SDSS}{{\Sp\Dp\Sp\Sp}}
\newcommand{\SSDS}{{\Sp\Sp\Dp\Sp}}

\newcommand{\SSSD}{{\Sp\Sp\Sp\Dp}}